\documentclass{cernyrep} 
\usepackage{texnames}
\usepackage[T1]{fontenc}
\newcommand{\be}{\begin{equation}} \newcommand{\ee}{ \end{equation}}
\newcommand{\ba}{\begin{eqnarray}} \newcommand{\ea}{ \end{eqnarray}}

\begin{document}
\title{Neutron width statistics using a realistic description of the neutron channel}
 % Neutron width statistics in a realistic statistical-reaction model
\author{ P.~Fanto$^{1}$, G.~F.~Bertsch$^{2}$, and Y.~Alhassid$^{1}$}
\institute{$^{1}$Center for Theoretical Physics, Sloane Physics Laboratory, Yale University, New Haven, Connecticut 06520, USA \\
$^{2}$Department of Physics and Institute for Nuclear Theory, 
Box 351560, University of Washington, Seattle, Washington 98195, USA\\
}

\maketitle % this produces the title block

\begin{abstract}
A basic prediction of the statistical model of compound nucleus reactions is that the partial widths for decay into any open channel channel fluctuate according to the Porter-Thomas distribution (PTD).  A recent experiment on $s$- and $p$-wave neutron scattering from platinum isotopes found that the experimental $s$-wave partial neutron width distributions deviated substantially from the PTD.  Several explanations for this finding have been proposed within the statistical model, but none has resolved this issue. Here, we review the application of a realistic resonance-reaction model to $s$-wave neutron scattering from $^{194}$Pt.  Our main conclusion is that the PTD provides an excellent description of the partial neutron width distribution, provided that the secular energy dependence of the average neutron width is correctly described.  Within a realistic range of model parameters, there can be a near-threshold bound or virtual state of the neutron channel that changes this secular dependence from the usual $\sqrt{E}$ dependence, as proposed by Weidenmüller~\cite{Weidenmuller2010}.  In this case, the use of the $\sqrt{E}$ dependence to analyze the data will lead to apparent deviations from the PTD.  We discuss the limited parameter range where such a near threshold state can have a significant effect.
\end{abstract}

\section{Introduction}\label{sec:intro}
Compound nucleus (CN) reactions are important in basic nuclear science, nuclear astrophysics, and nuclear technology applications.  %The CN describes an equilibrated intermediate state of the fused projectile and target nuclei.
% that survives for a long time, relative to the time the projectile would take to cross the nuclear volume.  
%Because the number of CN states at the neutron threshold energy is enormous, a statistical description of the CN is necessary.  
The statistical model of CN reactions~\cite{Mitchell2010} provides the theoretical framework for understanding this class of nuclear reactions.  The basic assumption of the statistical model is that the CN can be described by a random matrix drawn from the Gaussian orthogonal ensemble (GOE)~\cite{Mitchell2010,Alhassid2000}.  %The model consequently predicts fluctuations of the resonance energies and widths that resemble those observed in any chaotic quantum system that is invariant under time reversal.  
Energy-averaged cross sections and resonance parameters are calculated as averages over the ensemble.  The statistical model modifies the traditional Hauser-Feshbach theory of CN reactions~\cite{Hauser1952} %, which is based on Bohr's hypothesis of the independence of the CN formation and decay~\cite{Bohr1936}, 
through the so-called width fluctuation correction (WFC)~\cite{Kawano2015}.  The Hauser-Feshbach plus WFC approach is used in  statistical reaction codes~\cite{Koning2012}. 
%and EMPIRE \cite{Herman2007}. 

A basic prediction of the statistical model is that, for isolated resonances, the partial widths for any reaction channel fluctuate according to the Porter-Thomas distribution (PTD)~\cite{Porter1956, Mitchell2010}.  However, a relatively recent experiment~\cite{Koehler2010} found that the distributions of partial neutron widths in $s$-wave scattering of neutrons off $^{192,194,196}$Pt were significantly broader than the PTD.  A number of theoretical studies attempted to explain these deviations from PTD within the statistical model framework.  These studies proposed the following explanations:

(i).~In the experiment of Ref.~\cite{Koehler2010}, the resonance energies and partial widths were extracted using an $R$-matrix fit to measured data.  To obtain the partial width fluctuations, the partial widths must be divided by the secular dependence of the average neutron width on the resonance energy.  This energy dependence is expected to be proportional 
%to the squared penetrability within the $R$-matrix formalism, i.e., 
to $\sqrt{E}$ for $s$-wave neutrons.  Weidenm\"uller proposed~\cite{Weidenmuller2010} that, for the Pt isotopes, a near threshold bound or virtual state of the $s$-wave neutron channel\footnote{A bound state of the $s$-wave neutron channel corresponds to a pole of the $S$-matrix on the positive imaginary axis in the complex wavenumber $k$ space. As the potential is made less attractive, this pole crosses onto the negative imaginary axis and becomes a virtual state with negative energy and zero width.  See Ref.~\cite{Taylor1972} for details.} changes this secular energy dependence.  Thus, the use of the $\sqrt{E}$ form in the experimental analysis might have caused the observed PTD violation in Ref.~\cite{Koehler2010}.  

(ii).~Coupling to the reaction channels induces non-statistical interactions among the CN resonances.  These interactions consist of a purely imaginary term due to on-shell coupling to the channels, and a real term called the Thomas-Ehrman shift due to off-shell processes~\cite{Mitchell2010}.  In the regime of isolated resonances, these interactions are expected to be negligibly weak.  However, the experimental finding of Ref.~\cite{Koehler2010} motivated closer studies of the effect of these interactions on the fluctuations of the partial widths. It was shown numerically~\cite{Celardo2011} and analytically~\cite{Fyodorov2015} that even a relatively weak imaginary term could cause the partial width fluctuations to deviate from the PTD.  However, it is unclear if the coupling for isolated resonances is strong enough to cause this effect.  In Ref.~\cite{Volya2015}, it was proposed that the Thomas-Ehrman shift could be quite strong near the neutron threshold for the Pt isotopes and cause deviations from the PTD.  %Moreover, numerical tests of Ref.~\cite{Volya2015} seemed to show such an effect.  
However, Bogomolny subsequently proved that the only effect of an energy-independent real shift is to modify the secular variation of the average partial neutron width with energy~\cite{Bogomolny2017}.  %As this variation is always divided out to obtain the fluctuations, such a real shift cannot affect the PTD.  Moreover, this secular variation is slow, so the PTD survives when only a few resonances in the center of the spectrum are studied.  
The numerical results of Ref.~\cite{Volya2015} were obtained using the entire resonance spectrum without dividing out this secular variation.

%We note that the PTD violation has also been attributed to many-body correlations beyond the statistical model \cite{Volya2011}.  An accurate test of such correlations is difficult due to the huge model space of allowed states and thus has not been done. 

However, explanations (i) and (ii) have not been studied within a model that describes fully %includes all aspects of 
the coupling between the entrance neutron channel and the CN states.  Specifically, near threshold, the non-statistical interactions vary significantly with energy, and this variation has been neglected in the above studies.  %Moreover, investigating explanation %the effect of the near threshold state proposed in 
%(i) requires a realistic description of the entrance neutron channel, which has not previously been done.

Here, we review a recent study~\cite{Fanto2018} of $s$-wave neutron scattering from $^{194}$Pt within a model that combines a realistic description of the entrance neutron channel with the usual GOE description of the internal CN states.  %This model naturally includes all physical aspects of the coupling between the neutron channel and the CN.  Moreover, it enables the calculation of cross sections within the same framework in which resonance energies and widths are calculated.  
This model enables the calculation of cross sections and resonance energies and widths within the same framework.  We used a baseline set of the model parameters from the published literature. We then varied these parameters to investigate explanations (i) and (ii) above.  Our main conclusion is that the PTD provides an excellent description of the partial neutron width distribution, provided that the secular dependence of the average width on energy is correctly described.  Within our parameter range, there can be a near-threshold bound or virtual state of the neutron channel that modifies the usual $\sqrt{E}$ secular variation of the average neutron width.  In this case, use of the $\sqrt{E}$ form to analyze the width fluctuations yields apparent PTD violation.  We find that this effect is significant only within a limited range of the model parameters.  

\section{Resonance-reaction model}\label{sec:model}

We describe the $s$-wave neutron channel on a spatial mesh with spacing $\Delta r$ and $N_n$ radial sites $r_i = i\Delta r$ $(i = 1,...,N_n)$.  The neutron channel Hamiltonian matrix is obtained by discretizing the usual radial Schr\"odinger equation
\be\label{eq:H-neutron}
\mathbf{H_n}_{,ij} = [2t + V(r_i)]\delta_{ij} - t \delta_{i,j+1} - t\delta_{i,j-1} \, , \qquad i,j = 1,...,N_n\;,
\ee
where $t = \hbar^2/(2 \mu\Delta r^2)$ ($\mu$ is the reduced mass) and $V(r)$ is the neutron channel potential.  The neutron channel potential has a Woods-Saxon form $V(r) = -V_0 [1 + \exp((r-R)/a)]^{-1}$ where $R = r_0 A^{1/3}$ for mass number $A$ of the target nucleus.  The model includes $N_c$ internal CN states, taken from the middle third of the spectrum of a GOE matrix.  To account for gamma decay of the resonances, we add to each energy a constant imaginary term that corresponds to the total gamma decay width $\Gamma_\gamma$.  This gives the CN Hamiltonian matrix
\be\label{eq:H-CN}
\mathbf{H_c}_{,\mu\nu} = \delta_{\mu\nu} \left(\epsilon_\mu - \frac{i}{2}\Gamma_\gamma \right) \, , \qquad \mu,\nu = 1,...,N_c\;,
\ee
%where $\Gamma_\gamma$ is the total gamma width.  
The neutron channel is coupled to the internal states at one spatial site $r_{\rm ent} = i_{\rm ent}\Delta r$ by the coupling matrix
\be\label{eq:V}
\mathbf{V}_{i\mu} = \delta_{i, i_{\rm ent}} v_0 s_\mu(\Delta r)^{-1/2}\;.
\ee
In Eq.~(\ref{eq:V}), $v_0$ is a model parameter, $s_\mu$ is a normally distributed random variable accounting for the fluctuations of the GOE internal states, and the explicit $\Delta r$ dependence allows the results to be consistent in the continuum limit $\Delta r \to 0$.  The full Hamiltonian matrix $\mathbf{H}$ of dimension $(N_n+N_c)\times(N_n+N_c)$ is obtained by combining Eqs.~(\ref{eq:H-neutron}-\ref{eq:V}).
%\be\label{eq:H}
%\mathbf{H} = \left[ \begin{matrix} \mathbf{H_n} & \mathbf{V} \\ \mathbf{V}^T & \mathbf{H_c}\end{matrix}\right]
%\ee
%To calculate either resonances or cross sections, one combines the radial Schr\"odinger equation with the regularity condition $u(0) = 0$ and the appropriate asymptotic boundary condition.  This approach was introduced by G.~F. Bertsch in his computer program Mazama (see, e.g. Ref..~\cite{Bertsch2018}).

%\subsection{Resonance determination}
%To calculate resonances, or $S$-matrix poles, one considers the asymptotic form of the neutron wavefunction at the last mesh point $N_n$ and the point just beyond the edge of the mesh $N_{n+1}$.  
For a pole of the $S$-matrix, the wavefunction is asymptotically purely outgoing with a complex wavenumber $ k$, i.e,  $u(r) \to B( k) e^{i k r}$ for large $r$.  This form implies that, at the edge of the mesh,
\be
u(N_{n+1}) = u(N_n) e^{ik\Delta r}\;.
\ee
The Schr\"odinger equation consequently yields
\be\label{eq:poles}
\left(\mathbf{H} - te^{i k \Delta r} \mathbf{C} - \frac{\hbar^2 k^2}{2\mu}\right)\vec u = 0\;,
\ee
where ${\bf C}_{i j} = \delta_{i j}\delta_{i N_n}$ and $\vec u$ is an $(N_n+ N_c)$-component vector that represents the wavefunction in the model space.  We solve Eq.~(\ref{eq:poles}) iteratively to obtain the complex resonance wavenumbers $k_r$.  The resonance energies and total widths are then determined from $E_r - (i/2)\Gamma_r = \hbar^2 k_r^2/2\mu$, and the partial neutron widths are found from $\Gamma_{n,r} = \Gamma_r - \Gamma_\gamma$.  All calculations shown here used $(\Delta r, N_n, N_c) = (0.01\text{ fm}, 1500, 360)$.  More details can be found in the Supplemental Material of Ref.~\cite{Fanto2018}.

To calculate cross sections, one uses instead the asymptotic boundary condition $u(r) \to e^{-ikr} - S_{nn}(k) e^{ikr}$, where $S_{nn}(k)$ is the elastic neutron scattering amplitude (see Supplemental Material of Ref.~\cite{Fanto2018}).  The model discussed here was introduced by G.~F. Bertsch in his computer code Mazama~\cite{Bertsch2018}.

\section{Results}\label{sec:results}

\subsection{Model parameter sets}

The model described above has the following physical parameters: the parameters $(V_0, r_0, a)$ of the Woods-Saxon neutron channel potential, the average resonance spacing $D$ and their total gamma decay width $\Gamma_\gamma$, and the channel-CN coupling parameter $v_0$.  For the $n + ^{194}$Pt reaction, we used $(r_0,a) = (1.27,0.67)$ fm from Ref.~\cite{Bohr1969}, and $D = 82$ eV, $\Gamma_\gamma = 72$ meV from the RIPL-3 database~\cite{ripl}.  %These parameters were fixed in all our calculations.

%We varied the neutron channel potential depth $V_0$ and the average coupling parameter $v_0$.  %The former controls the presence of the near-threshold state crucial to explanation (i) of Sec.~\ref{sec:intro}.  The latter sets the strength of the non-statistical interactions and thus allows investigation of explanation (ii) of Sec.~\ref{sec:intro}.  
In Table~\ref{table:models}, we list the parameter sets $(V_0, v_0)$ used in our calculations.  The baseline set corresponds to $V_0 = -44.54$ MeV taken from Ref.~\cite{Bohr1969} and $v_0$ tuned to reproduce the RIPL-3 neutron strength function parameter $S_0 = 2 \times 10^{-4}$ eV$^{-1/2}$ at 8 keV of neutron energy (which is in the middle of the experimental range of Ref.~\cite{Koehler2010}).  In sets M2 and M3, we take $v_0$ to be, respectively, half and twice its value for the baseline set.  In sets M4--M6 with $V_0 = -41.15$ MeV,  a near-threshold bound state with energy $E_0 \approx -2$ keV exists in the neutron channel.  In M4  $v_0$ is tuned to reproduce the RIPL-3 neutron strength function parameter, while M5 and M6 correspond, respectively, to half and twice the value of $v_0$ for the M4 set.  %The values of $\chi^2$, $\nu_{\rm fit}$, and $\chi^2_{\rm fit}$ are discussed in more detail below.  
%Table \ref{table:models} also includes results from the calculations described in detail below.
\begin{table}[h!]
\centering
\begin{tabular}{| c  c c c c c c |}
\hline
 Model & Baseline & M2 & M3 & M4 & M5 & M6  \\ \hline
 $V_0$ (MeV) & -44.54 & -44.54 & -44.54 & -41.15 & -41.15 & -41.15 \\ 
 $v_0$ (keV-fm$^{1/2}$) & 11.0 & 5.5 & 22.0 & 1.6 & 0.8 & 3.2 \\ \hline
 $\overline \sigma_{\rm el}$ (b) & 30.  & 19.0  & 23. & 279. & 288.  &  249. \\
 $\overline \sigma_{\rm cap}$ (b) & 0.44 & 0.32 & 0.50 & 0.47 & 0.39 & 0.53 \\
 $\chi^2_r$ A & 0.9 & 1.0 & 1.1 & 0.9 & 1.0 &1.4 \\
 $\chi^2_r$ B & 1.0 & 1.0 & 1.3 & 5.8 & 6.0 & 6.1 \\
 \hline
 \end{tabular}
 \caption{\label{table:models} Various parameter sets used to study the $n + ^{194}$Pt reaction.  The average elastic scattering cross section $\overline \sigma_{\rm el}$ was evaluated at $E = 8$ keV, and the average neutron capture cross section $\overline \sigma_{\rm cap}$ was evaluated over the interval 5-7.5 keV.  The experimental value for  $\overline \sigma_{\rm cap}$ over this interval from Ref.~\cite{Koehler2013} is 0.6 b.  $\chi^2_r$ is the reduced $\chi$-squared parameter from comparing the reduced partial neutron width distributions with the PTD.  Reductions A and B are discussed in the text.  Adapted from Table 1 of Ref.~\cite{Fanto2018}.}
\end{table}

\subsection{Cross sections}

\begin{figure}[h!]
\centering
\includegraphics[width=0.5\textwidth]{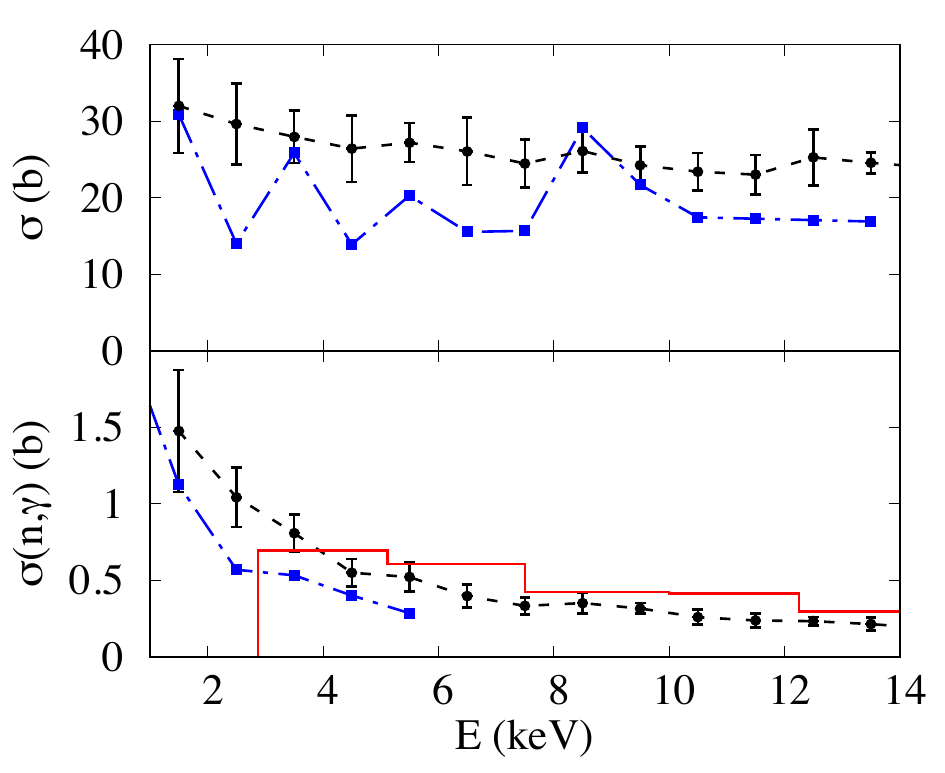}
\caption{\label{fig:xs-base} Elastic (top panel) and capture (bottom panel) cross sections.  Baseline model calculations (black circles connected by dashed lines) are compared with evaluated data from the JEFF-3.2 library (blue squares connected by dashed-dotted lines).  Error bars represent standard deviations from the 10 GOE realizations.  The red histogram is experimental data from Ref.~\cite{Koehler2013}.  Taken from Fig.~1 of Ref.~\cite{Fanto2018}.}
\end{figure}

In Fig.~\ref{fig:xs-base}, we show the cross sections for the baseline model, compared with evaluated data from the JEFF-3.2 database \cite{jeff} and experimental data from Ref.~\cite{Koehler2013}.  For the model calculations, the cross sections were averaged over 10 realizations of the GOE and over energy bins of 1 keV.  The agreement between our model calculations and the available data is reasonably good.  

\begin{figure}[h!]
\centering
\includegraphics[width=0.5\textwidth]{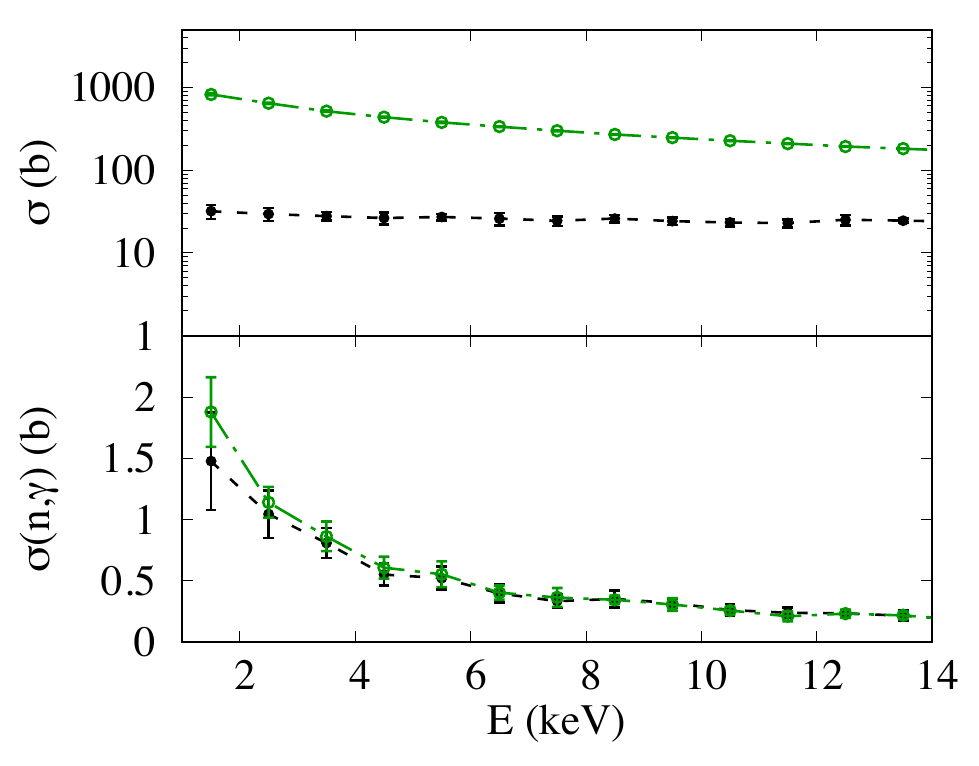}
\caption{\label{fig:xs-M4} Elastic (top panel) and capture (bottom panel) cross sections.  Baseline model calculations (black solid circles connected by dashed lines) are compared with calculations based on parameter set M4 of Table \ref{table:models} (green open circles connected by dashed-dotted lines).  The error bars represent standard deviations using 10 GOE realizations (see text).}
\end{figure}

In Fig.~\ref{fig:xs-M4}, we compare the average cross sections of the baseline parameter set and of the set M4 (see Table \ref{table:models}).  The presence of a near-threshold bound or virtual state in the neutron channel  enhances significantly the elastic scattering cross section, but does not affect much the capture cross section.  As is seen in Table~\ref{table:models}, the cross sections are not very sensitive to the coupling parameter $v_0$.

\subsection{Partial neutron width fluctuations}

\begin{figure}[h!]
\centering
\includegraphics[width=0.6\textwidth]{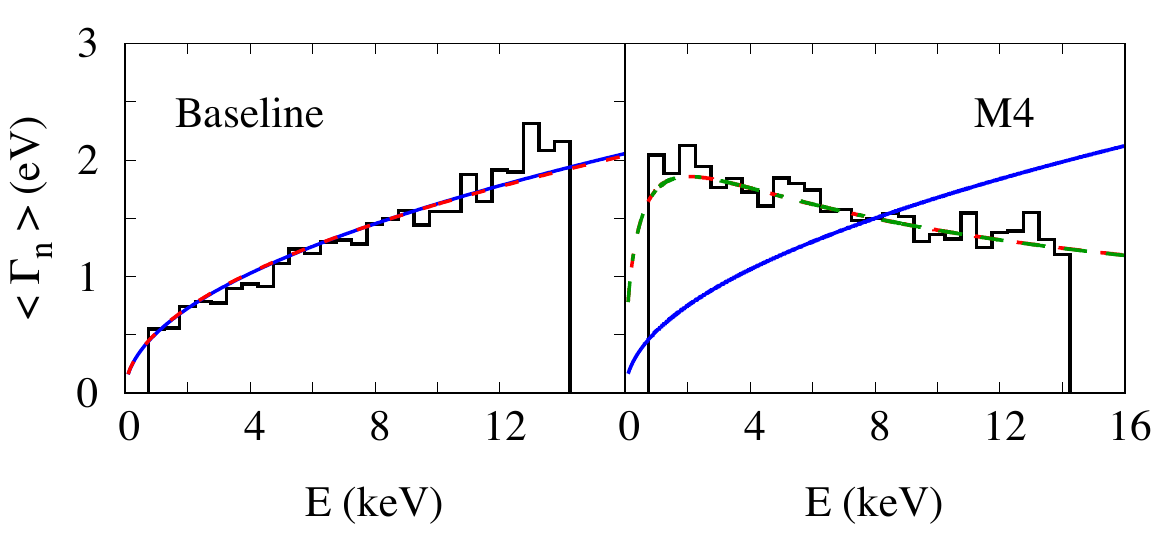}
\caption{\label{fig:average-width} Average neutron widths $\langle \Gamma_n\rangle$  as a function of neutron energy $E$.  The black solid histograms are the model calculations, the blue solid lines are proprotional to $\sqrt{E}$, and the red dashed lines are the squared neutron wavefunction $u^2_E(r_{\rm ent})$.  The green dashed-dotted line describes Eq.~(\ref{eq:wei}) with $E_0 \approx -2$ keV.  See text for further details.}
\end{figure}

For each parameter set in Table~\ref{table:models}, we calculated a set of resonance energies and partial neutron widths from 100 GOE realizations, taking from each realization 160 resonances whose real energies are at the center of the resonance spectrum.  This restriction to the center is necessary to avoid edge effects due to the finite number of resonances in the model.  In Fig.~\ref{fig:average-width}, we show the average partial neutron width $\langle \Gamma_n \rangle$ as a function of the neutron energy for the baseline and M4 sets.  We compare with the $\sqrt{E}$ dependence, as well as with the squared neutron wavefunction at the entrance point $u^2_E(r_{\rm ent})$.  This latter quantity is expected to describe the energy dependence of the average width within the statistical theory. In the right-hand panel, we also compare the formula derived by Weidenm\"uller~\cite{Weidenmuller2010}
\be\label{eq:wei}
\langle \Gamma_n \rangle (E) \propto \frac{\sqrt{E}}{E + | E_0 |} \;,
\ee
where $E_0$ is the energy of the near-threshold bound or virtual state.  This formula is in excellent agreement with our model calculations using $E_0 \approx -2$ keV.  We find similar results to those shown in the left and right panels of Fig.~\ref{fig:average-width} for other parameter sets of Table~\ref{table:models} with $V_0 = -44.54$ MeV and $V_0 = -41.15$ MeV, respectively.

\begin{figure}[h!]
\centering
\includegraphics[width=0.5\textwidth]{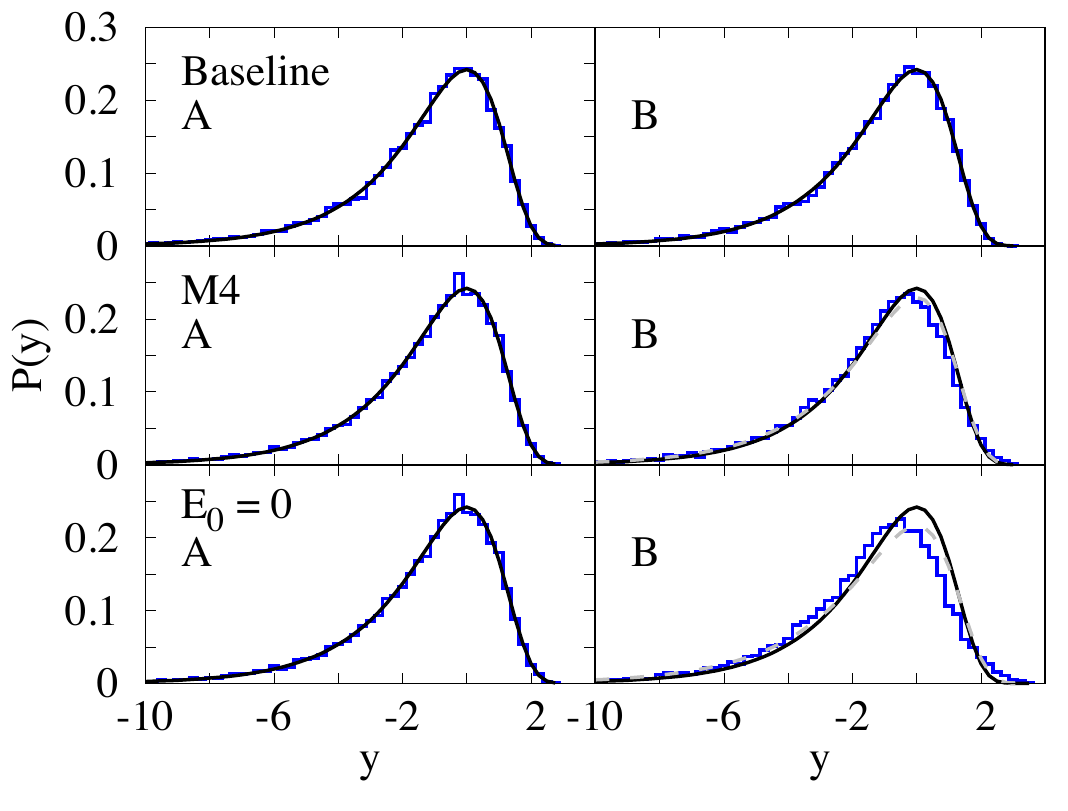}
\caption{\label{fig:dists} Reduced partial neutron width distributions calculated from the model (blue histograms) are compared with the PTD (black solid lines) for various parameter sets.  The grey dashed lines are fits of a $\chi$-squared distribution in $\nu$ degrees of freedom to the model results.  Shown are results for reductions A (left panels) and B (right panels). These reductions are discussed in the text.}
\end{figure}

The reduced partial neutron width for each resonance $r$ is defined as
\be\label{eq:red-widths}
\hat \Gamma_{n,r} = \Gamma_{n,r}/\langle \Gamma_{n}\rangle(E_r)\;.
\ee
According to the statistical theory, the fluctuations of the normalized reduced width $x = \hat \Gamma_n / \langle \hat \Gamma_n\rangle$ should follow the PTD.  We consider the distribution $P(y)$  of $y = \ln x$.  The PTD for y is given by
\be
\mathcal{P}_{\rm PTD} (y) = \sqrt{\frac{x}{2\pi}} e^{-x/2} \;.
\ee
We extracted the reduced widths (\ref{eq:red-widths}) in two ways.  In reduction A, we used the average widths calculated in the model as shown in Fig.~\ref{fig:average-width}, while in reduction B, we used the ansatz $\langle \Gamma_n \rangle (E)\propto \sqrt{E}$.  In Fig.~\ref{fig:dists}, we compare the distributions obtained using these two reductions with the PTD for various parameter sets.  For the baseline case, excellent agreement is obtained in both reductions A and B.  For set M4, in which there is a near-threshold state in the neutron channel, the reduced width distribution obtained with reduction B is broader than the PTD.  For a goodness-of-fit test, we calculated the reduced $\chi$-squared value $\chi^2_r$~\cite{chi-squared}, using $\chi^2_r \approx 1$ as the metric for a good fit.  The results in Fig.~\ref{fig:dists} are consistent with the $\chi^2_r$ values shown in Table~\ref{table:models}.  For set M4, we also show a best-fit $\chi$-squared distribution in $\nu$ degrees of freedom. The PTD corresponds to $\nu = 1$.  For set M4 in reduction B, we find $\nu_{\rm fit} = 0.92$, and the fitted distribution differs noticeably from the model results.  Similar results were found for different values of the coupling parameter $v_0$.  This parameter controls the strength of the non-statistical interactions among the resonances.  Thus, within our parameter range, these interactions do not significantly affect the partial neutron width fluctuations.

Eq.~(\ref{eq:wei}) indicates that the deviation between the correct average width energy dependence and the $\sqrt{E}$ form will be largest for $E_0 \approx 0$.  In our model, this occurs for $V_0 = -41$ MeV.  The bottom panel of Fig.~\ref{fig:dists} shows results from the parameter set $(V_0, v_0) = (-41\text{ MeV}, 1.4\text{ keV-fm$^{1/2}$}$).  The reduced width distribution obtained with reduction B is noticeably broader than in set M4.

\section{Conclusions}\label{sec:conclusions}

We have studied the partial neutron width statistics for the $n + ^{194}$Pt reaction within a model that combines a realistic description of the neutron channel with the GOE description of the CN.  Our main conclusion is that the PTD describes the reduced width distributions for a reasonably large range of physical model parameters, provided the energy dependence of the average width is correctly described.  Thus, explanation (ii) of Sec.~\ref{sec:intro} is ruled out.  Within our parameter space, there can be a near-threshold bound or virtual state of the neutron channel that modifies the energy dependence of the average neutron width from the $\sqrt{E}$ dependence used in the experimental data analysis, as proposed by Weidenm\"uller~\cite{Weidenmuller2010}.  If the $\sqrt{E}$ dependence is used to extract the reduced widths in such a case, broader distributions than the PTD are obtained.  

In order to explain the findings of Ref.~\cite{Koehler2010}, the bound or virtual state must have an energy of only a few keV from threshold in all three platinum  isotopes studied.  The experimentalists found that using Eq.~(\ref{eq:wei}) in their analysis did not improve the agreement between their data and the PTD~\cite{comment}.  However, they note that their data fitting might break down in the presence of such a state.  As shown in Fig.~\ref{fig:xs-M4}, the existence of this state enhances significantly the elastic scattering cross section.  A measurement of this cross section could thus shed light on the possible existence of a near threshold bound or virtual state. 
%could therefore confirm or fully rule out explanation (i) of Sec.~\ref{sec:intro}.  

%In conclusion, the experiment of Ref.~\cite{Koehler2010} poses a serious challenge to the statistical model.  Nearly every theoretical explanation for it within the statistical model framework has been ruled out.  More experimental work investigating breakdowns of the PTD in neutron widths would be useful.

\section*{Acknowledgements}
This work was supported in part by the U.S. DOE grant
Nos.~DE-FG02-00ER411132 and DE-FG02-91ER40608, and by the DOE NNSA Stewardship Science Graduate Fellowship under cooperative agreement No.~DE-NA0003864.  
We would like to thank  H.~A. Weidenm\"uller for useful discussions. PF and YA acknowledge the hospitality of the Institute for Nuclear Theory at the University of Washington, where part of this work was completed during the program INT-17-1a, ``Toward Predictive Theories of Nuclear Reactions Across the Isotopic Chart.''  This work was supported by the HPC facilities operated by, and the staff of, the Yale Center for Research Computing.

\end{document}